\documentclass[runningheads]{llncs}
\usepackage{graphicx}
\usepackage{xcolor}
\usepackage{enumitem} 

\usepackage{xcolor,comment}

\newlist{inlineroman}{enumerate*}{1}
\setlist[inlineroman]{itemjoin*={{; and }},afterlabel=~,label=(\roman*)}

\begin{document}

\raggedbottom
\title{Semantic Technology based Usage Control for Decentralized Systems}
%
%
\author{Ines Akaichi\orcidID{0000-0002-6020-5572}}
\authorrunning{Ines Akaichi}
%
\institute{Institute for Information Systems \& New Media,\\ Vienna University of Economics and Business\\
\email{ines.akaichi@wu.ac.at}\\
}
\maketitle              
\begin{abstract}
The sharing of data and digital assets in a decentralized settling is associated with various legislative challenges, including, but not limited to, the need to adhere to legal requirements with respect to privacy (e.g. \emph {data protection legislation}) and copyright (e.g. \emph {copyright legislation}). In order to enable software platform providers to manage data and digital assets appropriately and to provide more control to data and digital asset owners, usage control technologies could be used to make sure that consumers handle data according to privacy preferences, licenses, regulatory requirements, among others. In this research proposal, we explore the application of usage control in decentralized environments. In particular, we address the challenges related to the specification of usage control policies, the enforcement of the respective policies, and the usability of the tools that are used to administer them.
\end{abstract}
\section{Introduction} 
Modern decentralized systems, such as the \emph{Internet of Things (IoT)}, \emph{virtual data spaces}, and \emph{distributed knowledge graph applications} face a variety of challenges from a data and digital asset management perspective. According to Pretschner \cite{Pretschner2009ANOO},
data owners are reluctant to share their data with decentralized systems, as often they have no control over how their data are used. Additionally, Park and Sandhu \cite{10.1145/984334.984339} highlight that the sharing of data in decentralized environments goes beyond traditional access control, as existing solutions do not provide control over data usage once access to the data has been granted. Technologies that aim to address this challenge, which are usually classified as usage control or policy-based usage control, aim to ensure that data consumers handle data according to usage policies stipulated by data owners. Generally speaking, usage control is a generic term for data management software that supports data protection, copyright, and/or various legislative and institutional policies in a variety of domains, including, but not limited to, mobile software, cloud computing, industry 4.0, IoT, and collaborative software.
\begin{description}
\item [Problem statement.]
In our proposal, we address the problem of policy specification, enforcement and administration in decentralized usage control. 
While the majority of usage control policy languages are built according to domain-specific requirements, it is unclear whether existing domain/use case-specific proposals could be used for usage control in the general sense, where a single system may need to support privacy preferences, regulatory requirements, licensing, among others. Moving to the semantic web community, researchers have proposed various general-purpose policy languages that have not previously been explored in the context of usage control. Therefore, it is also unclear how these policy languages can be used to provide adequate support for the common structures encountered in usage control requirements.

\item [Contributions.]In this proposal, 
we seek to explore the use of semantic technologies, i.e., ontologies and underlying semantics, to develop a unified and flexible policy language that supports different types of usage control policies in various domains. Additionally, we plan to include an enforcement framework by using the semantics of our policy language in order to automatically check for policy adherence. Finally, we plan to demonstrate the suitability of our proposal by integrating our framework to the social linked data platform, SOLID\footnote{https://solidproject.org/}, which currently only supports access control.   

\item [Paper structure.]
The remainder of this research proposal is structured as follows: in Section 2, we present related work. In Section 3, we outline the working hypothesis that underlies our research proposal. Next, in Section 4, we present our progress made to date. In Section 5, we describe the methodology that guides our research aside to our work plan. Finally, we conclude our work in Section 6.

\end{description}
\section{Related Work}
The term \emph{usage control} was first introduced by Park and Sandhu \cite{10.1145/984334.984339} whose research focus on supporting the continuous monitoring of digital asset usage in dynamic distributed environments. Over the years, researchers have proposed various usage control conceptual models (cf. \cite{CAO2020998,10.1145/984334.984339}), policy languages and frameworks (cf. \cite{Jung2014,Schtte2018LUCONDF}). Other works focused on enforcing the respective policies, via proactive  or reactive mechanisms that aim to prevent security breaches and policy violations (cf. \cite{Jung2014,basin2011monpoly}). 

When it comes to the semantic web community, researchers proposed general policy languages and frameworks, such as KaoS \cite{Bradshaw1997KAoSTA}, Rei \cite{DBLP:conf/policy/KagalFJ03}, and Protune \cite{Bonatti2005DrivingAM} 
to govern and manage a range of constraints (e.g. access control, privacy preferences, regulatory constraints) that are encountered in a variety of distributed systems, such as \emph{multi-agent systems}, \emph{computing grids}, \emph{enterprise information systems}, and \emph{pervasive environments}.
More recent studies proposed policy languages tailored to support access control (cf. \cite{WebID2015,10.1145/2660517.2660530}), privacy preferences (cf. \cite{BonattiPrivacy2020,Garcia2008AWS}), licensing (cf. \cite{Pellegrini2019DALICCAL,ODRL2018}) and regulatory requirements (cf. \cite{BonattiPrivacy2020,DeVos2019ODRLPM}). 

In usage control, the majority of policy languages (cf. DUPO \cite{CAO2020998}, LUCON \cite{Schtte2018LUCONDF}) were developed according to domain-specific requirements in relation to \emph{mobile software, cloud computing, IoT, industry 4.0.}, \emph{networking, operating systems, and collaborative software}. Whereas, the policy languages that are meant to be \emph{domain-agnostic} are either not validated using use cases (e.g. OB-XACML \cite{kateb2014}) or are only evaluated in a specific domain (e.g. IND\textsuperscript{2}UCE \cite{Jung2014} and \cite{Silva2010}). Hence, it is unclear if the existing proposals could be used for usage control in the general sense, where a single system may need to support privacy preferences, regulatory requirements, licenses, among others. 
While on one hand, general semantic policy languages can be used to express a variety of constraints, it is also unclear how these policy languages can be used to provide adequate support for the common structures encountered in usage control requirements (e.g. normative rules, obligations bound to condition, system and environmental conditions, attributes update).
On the other hand, tailored policy languages are bound to the constraints for which they were developed and only cover requirements that are encountered in their respective areas. 

\section{Gap \& Hypothesis} 
Building on the existing challenge in the field of usage control, we see the need for a general policy language and framework that allows for the expression of different types of policies in usage control and is not tied to specific applications. 
As pointed out by Akaichi and Kirrane \cite{akaichi2022}, a usage control framework is a comprehensive framework that allows for the specification, enforcement and administration of usage policies. Accordingly, our framework has to incorporate the following key components: (i) a formal machine-readable policy language that is used to express usage control policies; (ii) an enforcement mechanism that can monitor compliance with said policies; and (iii) an administration interface that can be used to manage and monitor usage control policies.

Additionally, growing dynamic environments, such as the web or IoT-based data sharing systems, where new users continuously join, pose new challenges in terms of unpredictability and dynamicity. Therefore, decentralized environments bring an additional set of considerations from a usage control perspective with respect to: (i) controlling data that reside within multiple systems; (ii) securing data sharing and usage; and (iii) enforcing policies across multiple systems. As a result, the framework must also take into account the decentralized aspects of usage control.
To this end, we summarize the main hypothesis of our research proposal as follows:
\begin{quotation}
\textbf{Effective decentralized usage control may be achieved by: (i) a general-purpose policy language that can support different domains and applications of usage control; (ii) an enforcement mechanism that can address the challenges of dynamicity and unpredictability in decentralized environments
; and (iii) an administrative framework that offers users more control, trust and transparency over the use of their data.}
\end{quotation}
Our hypothesis leads to  the following research questions:\\\\
\emph{1) How can semantic web technologies be used to enhance the 
flexibility and extensibility of usage control policy languages?}\\
\emph{2)What are the most suitable mechanisms for enforcing usage control policies in decentralized environments?}\\
\emph{3) What are the most effective tools and techniques that can be used to provide data owners with more control, trust and transparency with respect to how their data are being used? }

\begin{figure}[t]
\includegraphics[width=9.5cm]{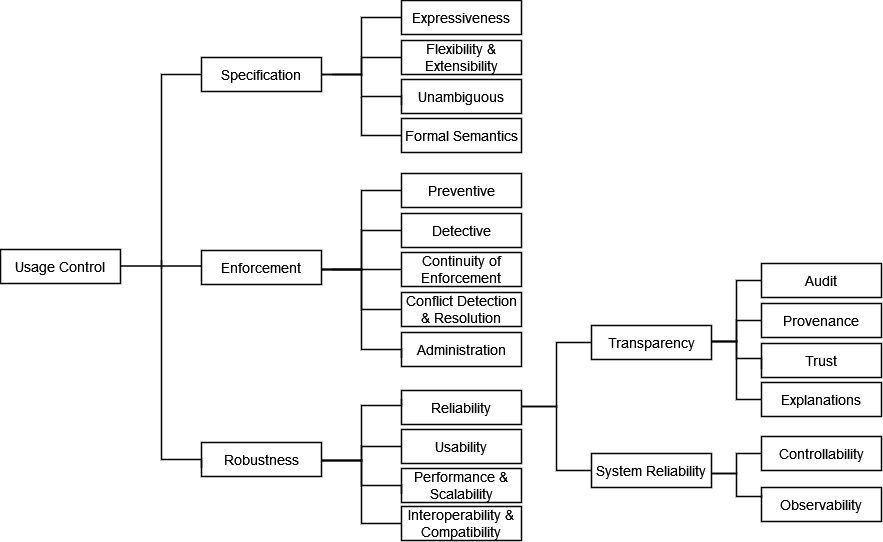}
\caption{Taxonomy of Usage Control Requirements}
\centering
\end{figure}
\section{Preliminary Results}
In an effort to establish an overview of what has been done in the field of usage control, a fundamental step was to gather and compare the predominant approaches to usage control, i.e., frameworks, found in the literature. To conduct this comparison, a very initial task was to examine the different requirements that have been used to guide the development of various usage control solutions. The requirements were then used to compare existing frameworks in order to assess their overall completeness. 

To this end, in our survey paper on usage control \cite{akaichi2022}, which is submitted to a Q1 journal and is currently under review,  we outline the following key contributions: \begin{inlineroman}
\item a taxonomy of usage control requirements brought from the literature. The taxonomy, which is depicted in Figure 1, is divided into three high level usage control dimensions, i.e. the specification of the policy language, the enforcement mechanism, and the robustness of the overall solution; \item the results of a qualitative comparison of the predominant usage control proposals \item various challenges and opportunities for the decentralized usage control domain that were derived from our comparison.
\end{inlineroman}


\section{Methodology \& Work Plan}
To answer our research questions, we adopt the design science research methodology (DSRM) presented by \cite{10.2753/MIS0742-1222240302}. Design science research is a paradigm focused on improving disciplinary knowledge based on the development of innovative artifacts. 
In Figure 2, we present our process model for conducting our research, which consists of the following activities:
\begin{figure}[t]
\includegraphics[width=11cm]{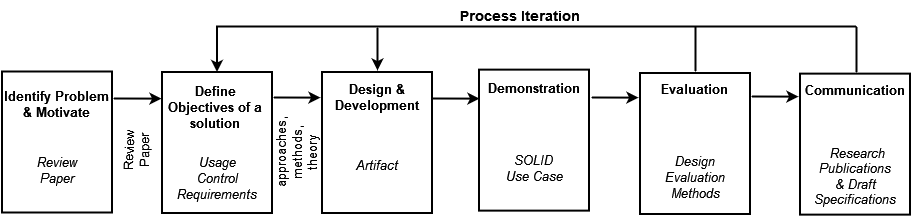}
\caption{Design Science Research Methodology Process Model adapted from \cite{10.2753/MIS0742-1222240302}}
\centering
\end{figure}
\paragraph{Identify Problem \& Motivate.} This activity defines the research motivation by pinpointing existing problems and gaps in a specific research area. To this end, our review article \cite{akaichi2022} outlined the state of the art in decentralized usage control and identified various gaps with respect to the specification, enforcement, and administration of policies.

\paragraph{Define Objectives of a Solution.} The objectives of a solution can be deduced from the problem definition. In our research, the objectives represent the requirements collected in the literature whereby a usage control solution is expected to address. Thus, we used the set of requirements to identify gaps in the domain by analyzing solutions and to what extent they cover these requirements. The full list of requirements is depicted in Figure 1. 

\paragraph{Design \& Development.} In this activity, artifacts are created. The challenges and opportunities presented in our overview paper drive the development of new artifacts, while the requirements determine the desired functionality of these artifacts. In our proposal, the following artifacts are to be considered:

\begin{description}
\item [A usage control policy language.]
We plan to develop a general purpose usage control policy language based on deontic operators with extended capability to include domain specific knowledge using semantic web technologies. Inspired by the SPECIAL policy language \cite{Bonatti2018SPECIAL}, we began developing the Usage Control Policy (UCP) language designed on the basis of deontic concepts (i.e., permission, prohibition, obligation, and dispensation) and constraints or conditions on data usage. In addition, the policy language is built on top of domain ontologies, which provides flexibility in expressing different types of usage control policies.
The initial version of our Usage Control Policy language only supports very simple conditions. Concretely, we plan to examine the suitability of various fine-grained conditions, such as actions that are bounded by cardinality restrictions and conditions that are tightly coupled to various actors and/or goals. Further, we want to study the expressiveness of various obligations and conditions and how they can be effectively structured into various policy profiles using Description Logic (DL) together with well-understood semantics and complexity. 
\item [An enforcement framework.] We plan to develop an enforcement framework that is able to leverage off the shelf reasoners, such as HermiT and FaCT++. Inspired by the works of \cite{DBLP:conf/policy/KagalFJ03,Bonatti2005DrivingAM}
, the DL based policy profiles together with the reasoning engine 
will be used to automatically check the compliance of data usage against usage control policies. To this end, we plan to leverage DL deductive reasoning capabilities to reason about usage control policies. As mentioned in \cite{BonattiPrivacy2020}, the advantage of using DL and consequently, OWL2, is that the majority of the policy-reasoning tasks are decidable and tractable, which is very important when making decisions regarding policy compliance in dynamic environments. 
In addition, we plan to explore the suitability of enforcement strategies (e.g. sticky policies \cite{Miorandi2020StickyPA}, logs \cite{Bonatti2017}, data flow tracking tools \cite{10.1145/3447867}) that enable  the enforcement of decentralized usage control.
\item [Data empowerment tools and technologies.]Empowering users means facilitating their awareness through tools that give data owners more control, trust and transparency over how their data are used. 
Thus, we plan to extend the existing SOLID administration application, which is initially used to manage access controls, by including usage control. 
The enhanced application aims to empower users by allowing them to share their preferences for how their data should be used, transparently display system-related decisions and actions in terms of how their data is actually used, and provide a secure and trusted environment for users to share their data, among other things.
To this end, we plan to explore various tools and techniques that can be used to provide data owners with more control, trust, and transparency, such as using transparency enhancement tools 
or trust management techniques. 
The development of this interface will be guided by various design principles emerging from the literature that are likely to support control, trust and transparency \cite{FischerHbner2016TransparencyPA}.
\end{description}

\paragraph{Demonstration.} 
This activity is used to demonstrate the effectiveness of artifacts in a given context that supports various policies such as access control, licensing, privacy, etc.  We plan to evaluate the suitability of the artifacts by extending the SOLID platform to support usage control, i.e., by integrating our usage control policy language and enforcement framework. In particular, we plan to apply the resulting platform to various use cases provided by the KnowGraphs\footnote{https://knowgraphs.eu/} project partners. For instance, the first use case originates from the IoT domain, in which it describes a data sharing platform that connects users to various IoT devices \cite{akaichi2022}. The second use case is from the financial domain, depicting a market data supply chain where different parties exchange data for financial instruments\footnote{https://w3c.github.io/market-data-odrl-profile/md-odrl-profile.html}. 

\paragraph{Evaluation.} 
This activity involves comparing the goals of a solution to the actual results observed when using the artifact in the demonstration. 
The evaluation of the adequacy of the policy language involves evaluating the expressiveness of the policy using the set of requirements that involve the specification dimension. While, the evaluation of the enforcement framework depends on using the set of requirements from the enforcement and robustness dimension. In turn, the evaluation of the administration framework depends on a couple of usability testing methods inspired by the work of legal and privacy researchers.
\paragraph{Communication.} This activity concerns communicating the results of our research. In our case, every artifact is mapped to a research article that will be submitted to journals and conferences, as well as to draft specifications.

\section{Conclusion} 
In this proposal, we explored the application of usage control in decentralized environments. Our gap analysis in the area of usage control identified various challenges in terms of specification, enforcement and administration of usage control policies. To this end, we discussed our research questions, our approach to addressing these challenges, our preliminary results, our future work, and the methodology that will guide our research.
\\
\\
\textbf{Acknowledgements.} This research is conducted under the supervision of Asst. Prof. Sabrina Kirrane and is funded by the European Union Horizon 2020 research and innovation
program under the Marie Sklodowska-Curie grant agreement No 860801.

%
%
%
\bibliographystyle{splncs04}
\bibliography{mybibliography}

\end{document}